\documentclass[reprint,amsmath,amssymb,prb,floatfix,twocolumn,showpacs]{revtex4}

\usepackage{graphicx}
\usepackage{dcolumn}
\usepackage{bm}
\usepackage{times}
\usepackage[usenames,dvipsnames]{color}
\usepackage{colordvi,epsf}
\usepackage{braket}
\usepackage{multirow}
\usepackage{longtable}
\usepackage[utf8]{inputenc}

\begin{document}

\title{Dzyaloshinskii-Moryia interaction at an antiferromagnetic interface:\\
 first-principles study of FeIr bilayers on Rh(001)}

\author{Sebastian Meyer}
\email[Email: ]{meyer@physik.uni-kiel.de}
\affiliation{Institut f\"ur Theoretische Physik und Astrophysik,
Christian-Albrechts-Universit\"at zu Kiel, D-24098 Kiel, Germany}

\author{Bertrand Dup\'e}
\altaffiliation[Present address: ] {Johannes Gutenberg-Universität Mainz,
Institute of Physics, Staudingerweg 7, D-55128 Mainz, Germany}
\affiliation{Institut f\"ur Theoretische Physik und Astrophysik,
Christian-Albrechts-Universit\"at zu Kiel, D-24098 Kiel, Germany}

\author{Paolo Ferriani}
\affiliation{Institut f\"ur Theoretische Physik und Astrophysik,
Christian-Albrechts-Universit\"at zu Kiel, D-24098 Kiel, Germany}

\author{Stefan Heinze}
\affiliation{Institut f\"ur Theoretische Physik und Astrophysik,
Christian-Albrechts-Universit\"at zu Kiel, D-24098 Kiel, Germany}

\date{\today}

\begin{abstract}
We study the magnetic interactions in atomic layers of Fe and $5d$ transition-metals such as Os, Ir, and Pt
on the (001) surface of Rh using first-principles calculations based on density functional theory. For both stackings of the $5d$-Fe
bilayer on Rh(001) we observe a transition from an antiferromagnetic to a ferromagnetic nearest-neighbor exchange interaction upon
$5d$ band filling. In the sandwich structure $5d$/Fe/Rh(001) the nearest neighbor exchange is significantly reduced.
For FeIr bilayers on Rh(001) we consider spin spiral states in order to determine exchange constants
beyond nearest neighbors. By including spin-orbit coupling we obtain the Dzyaloshinskii-Moriya interaction (DMI).
The magnetic interactions in Fe/Ir/Rh(001) are similar to those of Fe/Ir(001) for which an atomic scale spin lattice has been predicted.
However, small deviations between both systems remain due to the different lattice constants and the Rh vs.~Ir surface layers.
This leads to slightly different exchange constants and DMI and
the easy magnetization direction switches from out-of-plane for Fe/Ir(001) to in-plane for Fe/Ir/Rh(001). Therefore a fine tuning
of magnetic interactions is possible by using single $5d$ transition-metal layers which may allow to tailor antiferromagnetic skyrmions
in this type of ultrathin films.
In the sandwich structure Ir/Fe/Rh(001) we find a strong exchange frustration due to strong hybridization of the Fe layer with both
Ir and Rh which drastically reduces the nearest-neighbor exchange. The energy contribution from the DMI
becomes extremely large and DMI beyond nearest neighbors cannot be neglected. We attribute the large DMI to the low coordination of the Ir layer
at the surface.
We demonstrate that higher-order exchange interactions are significant in both systems which may be crucial for the magnetic ground state.
\end{abstract}

\maketitle

\section{Introduction}\label{Kap: Einleitung}

Magnetic skyrmions have been predicted in the late 1980's \cite{Bogdanov1989,Bogdanov1994},
but it took 20 years to confirm their existence experimentally \cite{Pfleiderer2009,Yu2010,Heinze2011,Romming2013}.
They have intriguing topological and dynamical properties which make them attractive for fundamental research 
and spintronic applications \cite{Kiselev2011, Fert2013, Nagaosa2013}.
After the first experimental observation of magnetic skyrmions in MnSi \cite{Pfleiderer2009},
they could be stabilized in different types of systems:
noncentrosymmetric bulk crystals \cite{Pfleiderer2009, Wilhelm2011, Munzer2010}, 
thin films of noncentrosymmetric crystals \cite{Tomura2012, Yu2011, Yu2010}
and ultrathin films \cite{Heinze2011, Romming2013}. 
The latter are composed of a few atomic transition-metal (TM) layers on surfaces.
Such systems have been studied extensively in the past decades,
since they are also at the heart of devices utilizing
the tunneling \cite{Julliere1975}
and the giant magnetoresistance \cite{Fert1988,Grunberg1989}.

A key ingredient for stabilizing skyrmions and other chiral magnetic configurations is the
Dzyaloshinskii-Moryia interaction (DMI) \cite{Dzyaloshinskii1957, Moriya1960} which
occurs due to spin-orbit coupling (SOC) in systems with broken inversion symmetry.
In 2007, the interfacial DMI due to the broken inversion symmetry at the surface \cite{Crepieux1998}
has been experimentally observed \cite{Bode2007} which opened the route to DMI stabilized 
skyrmions at interfaces such as the nanoskyrmion lattice of Fe/Ir(111) \cite{Heinze2011}.
An atomic adlayer of Pd changes the ground state of Fe/Ir(111) to a spin spiral which allows the creation
of isolated skyrmions in an applied magnetic field \cite{Romming2013,Dupe2014,Simon2014}. This demonstrates
the possibility of tailoring magnetic interactions in transition-metal films by changing the interface \cite{Dupe2014,Simon2014,Dupe2016,Rosza2016}.

Isolated skyrmions can be moved upon application of electric currents 
\cite{Jonietz1648, Schulz2012, Yu2012, Iwasaki2013, Fert2013, Lin2013, Iwasaki22013, Nagaosa2013, Zhang2015}.
However, skyrmions in materials with a ferromagnetic nearest-neighbor exchange interaction posses the disadvantage of being deflected
by the Magnus force \cite{Barker2016, Zhang2016}. This skyrmion Hall effect which has been recently observed in experiments
\cite{Jiang:17.1,Litzius:17.1} leads to skyrmion movement towards the edges of the tracks in sufficiently strong currents.
Skyrmions in antiferromagnets do not suffer from the Magnus force
because they have no net magnetization \cite{Barker2016, Zhang2016, Keesman2016, Jin2016}.
Therefore, they can be moved faster compared to ferromagnetic skyrmions.
However, so far there is no system in which these types of skyrmions have been observed.

Here, we study ultrathin film systems which combine antiferromagnetic nearest-neighbor (NN) exchange with large DMI
and are therefore potential candidates for skyrmions in antiferromagnets. 
We apply density functional theory (DFT) as implemented in the \textsf{FLEUR} code \cite{FLEUR}
and focus on atomic layers composed of Fe and a $5d$ transition-metal such as Os, Ir, or Pt on the Rh(001) surface.
We show that one atomic layer of the $5d$ element can change the magnetism of the system
from antiferromagnetic (Os) to ferromagnetic (Pt) similar as a 5\textit{d} surface \cite{Hardrat2009}.
The stacking of the bilayer has a large effect on the magnetism in the systems.
If the 5\textit{d} layer is the topmost layer, the NN exchange interaction decreases and
the systems are strongly exchange frustrated.

Bilayers of FeIr on Rh(001) are of particular interest since Rh and Ir are isoelectronic $4d$ and $5d$ 
transition-metals and have similar lattice constants.
It has been found before that the NN exchange is antiferromagnetic in both
Fe/Ir(001) \cite{Kudrnovsky2009, Deak2011, Kudrnovsky2013} and in Fe/Rh(001) \cite{Spisak2006, Al-Zubi2011}. 
For Fe/Ir(001) strong DMI has also been reported \cite{Belabbes2016, Hoffmann2015}
and the possible formation of an atomic spin lattice due to higher-order exchange interaction has been suggested \cite{Hoffmann2015}.
Antiferromagnetic exchange interactions at the interfaces of thin Fe films and multilayers on Ir(001) have also been 
observed experimentally \cite{Zakeri2013,Zakeri2014,Zakeri2017}.  
However, the Ir(001) surface exhibits a (\(5 \times 1\)) reconstruction which makes the preparation of a pseudomorphic Fe monolayer on Ir(001) 
difficult \cite{Schmidt2002, Hammer2003, Spisak2003}.
On the other hand, pseudomorphic growth of Fe on Rh(001) has been demonstrated experimentally and an antiferromagnetic checkerboard
ground state has been observed \cite{Kemmer2015} in agreement with theoretical predictions \cite{Spisak2006, Al-Zubi2011}.

For the two types of stackings of the FeIr bilayer -- Fe/Ir/Rh(001) and Ir/Fe/Rh(001) -- we obtain an 
antiferromagnetic NN exchange interaction. Exchange beyond nearest neighbors competes with the NN
interaction which leads to exchange frustration. In both systems we find a large DMI which induces
a spin spiral state. 
In Ir/Fe/Rh(001) the NN DMI even exceeds the NN Heisenberg exchange.
Upon introducing an additional Ir adlayer, however, 
the DMI is reduced by 50\% compared to Ir/Fe/Rh(001) leading to a collinear ground state.
We find that higher-order exchange interactions are significant for both bilayer stackings.
Our first-principles calculations show that FeIr bilayers on Rh(001) are promising candidates
for noncollinear spin structures with antiferromagnetic NN exchange such as isolated antiferromagnetic
skyrmions or antiferromagnetic skyrmion lattices as in Ref.~\onlinecite{Hoffmann2015}.

The paper is structured as follows: Section \ref{Kap: Methoden} describes the method 
and computational details of our calculations. In section \ref{Kap: Ergebnisse} we first
discuss the collinear states of the different Fe/5\textit{d} bilayers on Rh(001).
Afterwards, we show results of noncollinear calculations of
a freestanding Fe/Ir bilayer 
and we present the film systems Fe/Ir/Rh(001), Ir/Fe/Rh(001) and Ir/Ir/Fe/Rh(001).
We end with conclusions in section \ref{Kap: Vergleich}.

\begin{figure}
\includegraphics[scale=1]{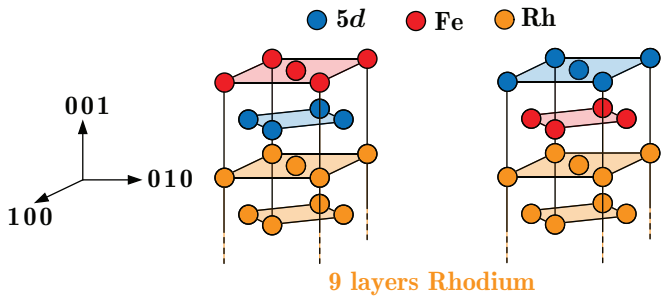}
\caption{(color online) Unit cell of Fe/5\textit{d} bilayers on Rh(001).
The 5\textit{d} elements are
Os, Ir, or Pt. Two different stackings of the bilayer are considered. Left:
the Fe layer at the surface. Right: the Fe layer in a sandwich structure between
the $5d$ layer and the Rh surface.
}
\label{Abb: Systeme}
\end{figure}

\section{Computational details}\label{Kap: Methoden}

We use the full-potential linearized augmented plane wave method
(FLAPW) \cite{Wimmer1981, Jansen1984} in film geometry \cite{Krakauer1979} as implemented in the J\"ulich DFT code 
\textsf{FLEUR}\cite{FLEUR}.
We performed spin-polarized calculations for every system
and we chose the same radii for the muffin-tin spheres for the three kind of
atoms (Fe: 2.26 a.u., Rh: 2.41 a.u., 5\textit{d}: 2.30 a.u.).
The lattice constant of our substrate (\(a = 3.84 \,\text{\AA} \)) 
was determined for Rh bulk within the generalized gradient approximation (GGA)
of the exchange-correlation (\textit{xc}) functional \cite{rPBE}.

\subsection{Structural relaxation}\label{Kap: Methoden-Relaxation}

For structural relaxations we used a symmetric film with 5 layers of Rh and a Fe/5\textit{d} bilayer
on both sides. We considered two types of stackings: Fe/5\textit{d} and 5\textit{d}/Fe (see Fig.~\ref{Abb: Systeme}).
We use the checkerboard c$(2\times 2)$ antiferromagnetic (AFM) state in the Fe layer and
minimize the forces between the uppermost layers in (001)-direction while three Rh layers are kept fixed \cite{Forces}.
We relaxed the structure with spin-polarized calculations using the GGA  of the
\textit{xc}-potential (revised Perdew-Burke-Ernzerhof, rPBE\cite{rPBE}).
The \textit{k}-point mesh consists of 136 \textit{k}-points in $\frac{1}{8}$ of the Brillouinzone (BZ)
and the cutoff for the basis functions is \(k_{max} = 5.0\, \text{a.u.}^{-1}\).
Relaxations were performed until the forces were less than $10^{-5}$~htr/a.u.
The equilibrium interlayer distances for all systems are given in table \ref{Tab: Relaxation}.
For the freestanding FeIr bilayer system we chose the in-plane lattice constant of Rh and
the layer distance according to Fe/Ir/Rh(001) (cf.~Tab.~\ref{Tab: Relaxation}).

\begin{table}
 \centering
 \caption {Interlayer distances in \(\text{\AA}\) after structural relaxation for
 the film systems Fe/5\textit{d}/Rh(001), 5\textit{d}/Fe/Rh(001)
 and Ir/Ir/Fe/Rh(001) in the c(\(2 \times 2\)) antiferromagnetic state
 as well as the chosen distance in the freestanding FeIr bilayer.
 Note that the last relaxed layer in the film systems is the Rh surface layer.
 A (\( -\)) indicates an interlayer distance according to the unrelaxed Rh(001) surface.}
\begin{ruledtabular}
 \begin{tabular}{lcccc}
& \(d_{12}\)  & \(d_{23}\) & \(d_{34}\) & \(d_{45}\)\\ \hline
Fe/Os/Rh(001) & 1.62 & 1.95 & 1.97 & \( - \)\\
Fe/Ir/Rh(001) & 1.69 & 2.02 & 1.91 & \( - \)\\
Fe/Pt/Rh(001) & 1.79 & 2.10 & 1.90 & \( - \)\\
Os/Fe/Rh(001) & 1.67 & 1.91 & 1.96 & \( - \)\\
Ir/Fe/Rh(001) & 1.68 & 1.89 & 1.96 & \( - \)\\
Pt/Fe/Rh(001) & 1.83 & 1.79 & 1.98 & \( - \)\\
Ir/Ir/Fe/Rh(001) & 2.11 & 1.71 & 1.87 & 1.96 \\ \hline
Fe/Ir & 1.69 & \(-\) & \(-\) & \(-\)
\end{tabular}
\end{ruledtabular}
\label{Tab: Relaxation}
\end{table}

\subsection{Collinear magnetic calculations}\label{Kap: Methoden-Kollinear}

In order to investigate the Fe/5\textit{d} bilayers on Rh(001) with respect to collinear magnetic order,
we use the optimized parameters of the structural relaxation to construct asymmetric films.
The setup is shown in Fig.~\ref{Abb: Systeme} where the surface is represented by nine layers of Rh(001).
The Fe/5\textit{d} bilayers are on one side of the substrate. 
We calculated the energy difference \( \Delta E\) between the ferromagnetic (FM)
and the c(\(2\times 2\)) AFM state in scalar-relativistic approximation \cite{Koelling1977}
using 484 \textit{k}-points in \(\frac{1}{4}\) of the BZ using the  
local density approximation (LDA) \cite{VWN}.
The cutoff for the basis functions was \(k_{max} = 4.0 \,\text{a.u.}^{-1}\).

\subsection{Spin-spiral calculations and Heisenberg exchange}\label{Kap: Methoden-Austausch}

To obtain the exchange constants $J_{ij}$ of the Heisenberg model for FeIr bilayers on Rh(001)
we calculate the energy dispersion of homogeneous, flat spin spirals \cite{Kurz2004, Zimmermann2014}.
These are characterized by their spin spiral vector $\mathbf{q}$ which
gives the propagation direction of the spiral. 
The $\mathbf{q}$ vector represents a vector in the reciprocal space and is chosen along
high symmetry directions of the BZ. 
The magnetic moment of atom $i$ is given by $\mathbf{M}_i = M (\cos{(\mathbf{q} \cdot \mathbf{R}_i)} \sin \theta, \sin{(\mathbf{q} \cdot \mathbf{R}_i)} \sin \theta,\cos \theta)$
where $\mathbf{R}_i$ is the position of atom $i$ and $\theta$ is the opening angle of the spiral. For the flat spirals considered here $\theta=90\,^\circ$.

In the absence of spin-orbit coupling the generalized Bloch theorem can be applied
to calculate spin spirals within the chemical unit cell of the system \cite{Sandratskii1991}.
Asymmetric films with 9 Rh substrate layers and the FeIr bilayer on one side 
as described in section \ref{Kap: Methoden-Kollinear} were used for the spin spiral calculations.
We apply the exchange-correlation functional in LDA \cite{VWN}
and a dense \textit{k}-point mesh of 48$\times$48 k-points
in the full two dimensional BZ.
The energy cutoff is set to \(k_{max} = 4.0\, \text{a.u.}^{-1}\).
The interlayer distances from the structural relaxation obtained for the c$(2 \times 2)$ AFM 
ground state is chosen (cf.~table \ref{Tab: Relaxation}).

The resulting energy dispersion curves \(E(\mathbf{q})\) along the \(\overline{X}-\overline{\Gamma}-\overline{M}\) direction
are mapped to the Heisenberg model 

\begin{equation}
 \mathcal{H} = - \sum_{ij} J_{ij} (\mathbf{m}_i \cdot \mathbf{m}_j) \,\text{.}\label{Gl: Heisenberg Modell}
\end{equation}

to obtain the shell resolved exchange constants $J_{ij}$ where $\mathbf{m}_i=\mathbf{M}_i/M_i$ is the unit vector
of the magnetic moment at atom site $i$.

\subsection{Dzyaloshinskii-Moriya interaction}\label{Kap: Methoden-DMI}

\begin{figure}
\includegraphics[scale=0.7]{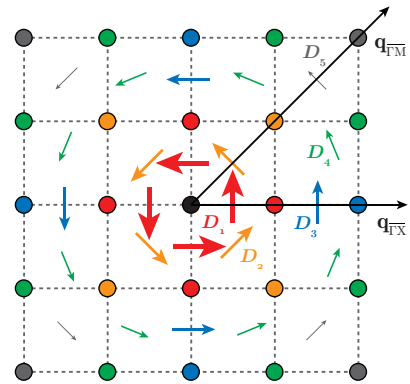}
\caption{Sketch of the Dzyaloshinskii-Moriya vectors for the Fe monolayer on the
Rh(001) surface from 1st to 5th neighbors (1st red, 2nd orange, 3rd blue, 4th green, 5th grey)
with the directions of the high symmetry lines of the 2 dimensional Brillouinzone.
The DM vectors are perpendicular to the bond between the black reference Fe atom 
and the corresponding neighbor. The size of the vectors illustrate the expected decreasing strength of the DMI with distance.
The propagation directions of spin spirals for $\mathbf{q}$ along the \(\overline{\Gamma}-\overline{M}\) and the \(\overline{\Gamma}-\overline{X}\) direction
are shown.
}
\label{Abb: Methoden-DMI}
\end{figure}

The degeneracy of the energies of left and right-rotating spin spirals
described above (Sec.~\ref{Kap: Methoden-Austausch}) is lifted
if spin-orbit coupling (SOC) is considered.
Two additional energy contributions will appear due to SOC:
the magnetocrystalline anisotropy (MAE, cf.~Sec.~\ref{Kap: Methoden-MAE})
and the antisymmetric exchange interaction,
the so called Dzyaloshinskii-Moriya interaction (DMI).
The latter requires a broken inversion symmetry,
which is given by the interfaces and surface in our systems.
The DMI can be described in the spin model via

\begin{equation}
 \mathcal{H}_{DMI}= - \sum_{ij} \mathbf{D}_{ij} \cdot 
 \left ( \mathbf{m}_i \times \mathbf{m}_j \right ) \text{,} \label{Gl: DMI}
\end{equation}

where \(\mathbf{D}_{ij}\) is the Dzyaloshinskii-Moriya (DM) vector
which determines the strength and the sign of the DMI.
Due to the cross product,
the DMI prefers a canting of magnetic moments \(i,j\) with one particular
rotation direction.
Typically, the DMI gives a small energy contribution compared
to the Heisenberg exchange.
The energy of DMI will be maximum for a rotation axis which
is parallel to the DM vector, which is shown in Fig.~\ref{Abb: Methoden-DMI}.
Therefore, we consider flat homogeneous cycloidal spin spirals.

If SOC is included, the generalized Bloch theorem is not valid anymore.
In principle, one can calculate spin spirals with SOC
in large supercells, however, the computational effort increases drastically.
Since SOC is typically a small effect one can treat it in first order perturbation theory \cite{Heide2009, Zimmermann2014}
starting from the self-consistent spin spiral calculations. 
The change of energy due to SOC is obtained from

\begin{equation}
 \Delta E_{\rm SOC} (\mathbf{q}) = \sum_{\mathbf{k},\nu} n_{\mathbf{k},\nu}(\mathbf{q}) \braket{\psi_{\mathbf{k},\nu}(\mathbf{q}) \vert \mathcal{H}_{SOC} \vert \psi_{\mathbf{k},\nu}(\mathbf{q})} \text{,} \label{Gl: 1. Ordnung SOC}
\end{equation}

where \(\mathcal{H}_{SOC}\) is the Hamilton operator of SOC,
\(\psi_{\mathbf{k},\nu}(\mathbf{q})\) is the selfconsistent wavefunction of the spin spiral state
and $n_{\mathbf{k},\nu}(\mathbf{q})$ is the weight of the state to the BZ summation.
Due to the symmetry of our ultrathin films at a surface the DM vectors lie in the surface plane as shown in Fig.~\ref{Abb: Methoden-DMI}
and therefore we consider cycloidal spin spirals.
In order to extract the strength of the DMI, Eq.~\eqref{Gl: DMI} is fitted to the SOC contribution of the system.
The cut-off parameters for the calculation of the SOC contribution in first-order
perturbation theory are identical to those from the spin spiral calculations (cf.~Sec.~\ref{Kap: Methoden-Austausch}).

To test the use of first order perturbation theory for SOC to determine the energy contribution due to DMI
(see Sec.~\ref{Kap: Methoden-DMI}), we can perform self-consistent calculations including SOC
for certain spin spiral states. Since the generalized Bloch theorem cannot be used,
we have to calculate the total energies in large unit cells corresponding to the spin spiral periods.
Due to the large computational effort we restrict these calculations to the freestanding FeIr bilayer, 
i.e.~without the Rh(001) surface. Each spin spiral state has to be calculated separately in the
2D unit cell corresponding to its periodicity. 
We apply SOC in \(x\)-direction to left and right rotating cycloidal spirals and
use the energy cutoff of \(k_{max}= 3.8\, \text{a.u.}^{-1} \)
with different \(k\)-point sets adopted to each spiral state/unit cell
in order to obtain the same k-point density as in the spin spiral calculations.

\subsection{Magnetocrystalline anisotropy}\label{Kap: Methoden-MAE}

The second effect due to spin-orbit coupling (SOC) is the magnetocrystalline anisotropy energy (MAE).
We perform self-consistent scalar-relativistic calculations and use the force theorem \cite{Oswald1985, Liechtenstein1987} 
to apply SOC in \(z\) and \(x\) direction using the 2nd variation method \cite{Li1990}. 
The difference of the resulting two energies is \(K = E_{\perp} - E_{\parallel}\).
For the FeIr bilayers on Rh(001) we perform the calculations for the checkerboard c\((2\times 2)\) AFM state
which has the lowest total energy of all considered collinear states.
Asymmetric films with 9 Rh substrate layers and the FeIr bilayer on one side 
as described in section \ref{Kap: Methoden-Kollinear} were used.
For all systems, we choose 2025 \(k\)-points in the full Brillouin zone (BZ) and \(k_{max} = 4.0\,\text{a.u.}^{-1}\).

\subsection{Higher-order exchange interactions} \label{Kap: Methoden-4Spin}

\begin{figure}
\includegraphics[scale=1]{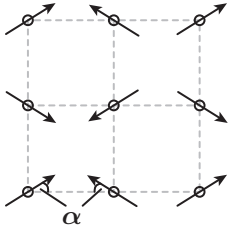}
\caption{Considered spin structure to test the influence of higher-order exchange interactions.
The angle \(\alpha\) is varied from 0 to \(45^\circ \) where theses structures correspond to the p\((2\times 1)\) antiferromagnetic state
and the 2Q state, respectively.}
\label{Abb: Winkel}
\end{figure}

The Heisenberg exchange interaction can be found as the second order expansion
in kinetic energy \cite{MacDonald1988} of the Hubbard model \cite{Hubbard1963}.
The fourth order gives rise to the 4-spin interaction and the biquadratic interaction.
The former can be understood as the hopping of electrons between four lattice sites, e.g. \(1 \rightarrow 2 \rightarrow 3 \rightarrow 4 \rightarrow 1\)
and is given by
\begin{align}
 \mathcal{H}_{4-spin} = - \sum_{ijkl} K_{ijkl} 
 \left [(\mathbf{m}_i\mathbf{m}_j) (\mathbf{m}_k\mathbf{m}_l) + \right. \\ \nonumber
 \left. (\mathbf{m}_j\mathbf{m}_k)(\mathbf{m}_l\mathbf{m}_i)
  -(\mathbf{m}_i\mathbf{m}_k) (\mathbf{m}_j\mathbf{m}_l) \right ] \,\text{.} \label{Gl: 4-Spin}
\end{align}
The biquadratic term arises due to the hopping of electrons between two sites
\(1 \rightarrow 2 \rightarrow 1 \rightarrow 2 \rightarrow 1\)
and is given by
\begin{equation}
 \mathcal{H}_{bi} = -\sum_{ij} B_{ij} \left ( \mathbf{m}_i \cdot \mathbf{m}_j \right )^2 \, \text{.}\label{Gl: biquadratische}
\end{equation}
\(K_{ijkl} \text{ and } B_{ij}\) depend on 
the electronic structure of the system similar to \(J_{ij}\) in Eq.~\eqref{Gl: Heisenberg Modell}.
Due to the perturbative expansion these higher-order exchange interactions 
are typically much smaller than the Heisenberg exchange and are often neglected.
To see the effect of the higher-order exchange in DFT it is in general necessary to consider two-dimensionally modulated noncollinear spin structures.
We compare states formed from superpositions of symmetry equivalent spin spirals. With respect to the Heisenberg exchange
these superpositions are degenerate with the spin spirals.
Energy differences obtained within a DFT calculation are therefore an indication of higher-order terms.

We choose the row-wise p\((2\times 1)\) antiferromagnetic state (cf.~Fig.~\ref{Abb: Winkel}, \(\alpha = 0^\circ\))
and change the angle \(\alpha \) of the spins up to \(45^\circ\) which
corresponds to the 2Q state \cite{Ferriani2007}. 
In nearest-neighbor approximation of the 4-spin and biquadratic interaction the energy as a function of $\alpha$ is 
given by $E(\alpha) = (2K_{\text{4-spin}} + B) \cdot \cos^2(2\alpha)$, i.e.~both terms possess the same angle 
dependence. Although one cannot obtain the two constants separately, these calculations allow us to estimate
the energy contributions from higher-order exchange interactions.

Asymmetric films as described in section \ref{Kap: Methoden-Kollinear} were used.
We apply LDA \cite{VWN} and a \textit{k}-point mesh of 576 k-points
in the full two dimensional BZ. The energy cutoff is set to \(k_{max} = 4.3\, \text{a.u.}^{-1}\).

\section{Results}\label{Kap: Ergebnisse}

\subsection{Collinear magnetic states of Fe/5\textit{d} bilayers on Rh(001)}

\begin{figure}
\includegraphics[scale=1]{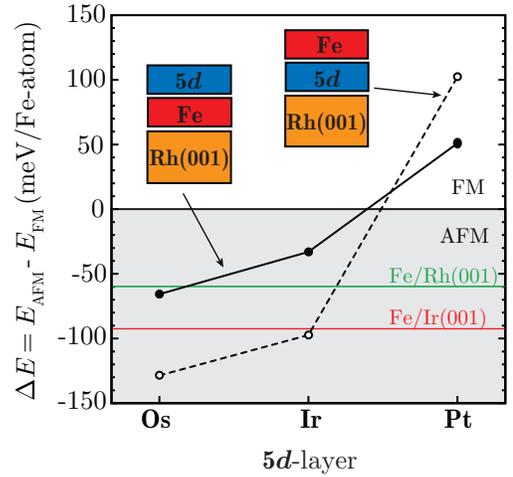}
\caption{(color online) Calculated total energy differences $\Delta E$ between the FM and the AFM state
for Fe/5\textit{d}/Rh(001) (dashed line) and 5\textit{d}/Fe/Rh(001) (solid line).
Positive values indicate that the FM state is preferred, negative values denote a favorable
$c(2\times 2)$ AFM structure.
All energies are calculated for structurally relaxed films in the AFM state.
The green (red) line is the value taken from Ref.~\onlinecite{Al-Zubi2011} (Ref.~\onlinecite{Hoffmann2015}).
}
\label{Abb: Kollineare_Energien}
\end{figure}

We start our study of the magnetic properties of Fe-$5d$ bilayers on Rh(001) by considering collinear magnetic states,
i.e.~the ferromagnetic (FM) and the c$(2 \times 2)$ antiferromagnetic (AFM) state.
Figure \ref{Abb: Kollineare_Energien} shows the total energy difference \(\Delta E = E_{AFM} - E_{FM}\)
for both stackings of Fe/5\textit{d} bilayers on Rh(001) and varying the $5d$ transition-metal from Os to Pt.
Negative energies indicate that the $c(2\times 2)$ AFM state
is favorable, positive values denote a preferred FM order.
The green and red line are two reference values from the literature for Fe/Rh(001) \cite{Al-Zubi2011}
and Fe/Ir(001) \cite{Hoffmann2015}. 

\begin{table}
 \centering
 \caption {Calculated magnetic moments for the upmost
 three layers in Fe/5\textit{d}/Rh(001) and 5\textit{d}/Fe/Rh(001) in
 \(\mu_B\) in the c(\(2 \times 2\)) antiferromagnetic (AFM)
 and the ferromagnetic (FM) state. 
 All calculations are performed in the structural relaxation of the AFM state.
 Note that in the c(\(2\times 2\)) AFM state the magnetic moments of adjacent layers 
 vanish due to symmetry.}
\begin{ruledtabular}
 \begin{tabular}{ccccccc}
 & \( \mu_{\text{Fe}}^{\text{AFM}}\) & \( \mu_{5d}^{\text{AFM}}\) & \( \mu_{\text{Rh(001)}}^{\text{AFM}}\) 
 & \( \mu_{\text{Fe}}^{\text{FM}}\) & \( \mu_{5d}^{\text{FM}}\) & \( \mu_{\text{Rh(001)}}^{\text{FM}}\) \\ \hline
Fe/Os/Rh(001) & 2.34 & 0.0 & 0.02 & 2.00 & \(-0.10\)& \(-0.08\) \\
Fe/Ir/Rh(001) & 2.71 & 0.0 & 0.10 & 2.67 & 0.10 & \(-0.14\)\\
Fe/Pt/Rh(001) & 2.95 & 0.0 & 0.13 & 3.01 & 0.25 & \(-0.02\)\\
Os/Fe/Rh(001) & 2.10 & 0.0 & 0.0 & 1.91 & \(-0.15\) & \(0.12\) \\
Ir/Fe/Rh(001) & 2.43 & 0.0 & 0.0 & 2.30 & 0.11 & 0.05 \\
Pt/Fe/Rh(001) & 2.83 & 0.0 & 0.0 & 2.81 & 0.29 & 0.13 \\
\end{tabular}
\end{ruledtabular}
\label{Tab: MM}
\end{table}

\begin{figure*}
\includegraphics[scale=1]{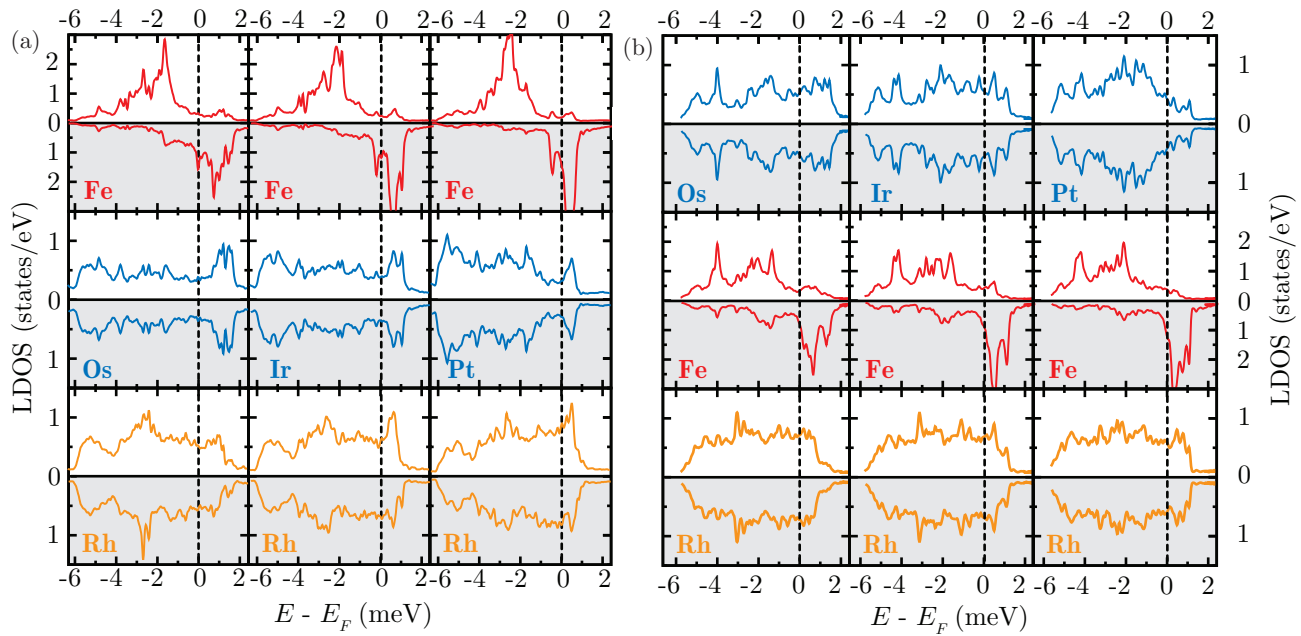}
\caption{(color online) Calculated spin-resolved local density of states (LDOS) of the top three layers of (a) Fe/5\textit{d}/Rh(001) and 
(b) 5\textit{d}/Fe/Rh(001) in the c(\(2 \times 2\)) antiferromagnetic state. Upper (lower) parts of each panel correspond to the majority 
(minority) spin channel.}
\label{Abb: LDOS}
\end{figure*}

First we focus on the bilayer stacking with the Fe layer at the surface.
There is transition in magnetic order from AFM to FM with the band filling of the 5\textit{d} layer.
This trend is similar to the one reported for Fe monolayers on 4\textit{d} and 5\textit{d} surfaces reported by Hardrat \textit{et al.} \cite{Hardrat2009} .
Note that we have chosen the relaxed geometry of the AFM state also to compute the total energy of the FM state to
be consistent with the spin spiral calculations in the following sections. 
However, using the structural relaxation of the FM state does not lead to
a qualitative change of the trend.
We conclude that already a single atomic layer of a $5d$ transition metal 
is sufficient to change the magnetic order in the Fe monolayer.
We attribute this finding to the fact that the 3\textit{d}-5\textit{d} hybridization 
which plays the key role for the change of the exchange interaction in the Fe layer
is an interface effect.
This interpretation is supported by the energy difference of Fe/Ir/Rh(001) being almost the same as that of Fe/Ir(001) \cite{Hoffmann2015}.

Upon changing the stacking of the Fe-5\textit{d} bilayer such that Fe is sandwiched between the $5d$ overlayer and the Rh(001) surface 
we observe a reduction of the energy difference.
Since the nearest-neighbor (NN) exchange interaction in the Fe layer is approximately proportional to the energy difference $\Delta E$,
this shows that \(J_1\) can be tuned by the stacking order of the Fe/5\textit{d} bilayer. In the following sections we will show for Ir 
as the $5d$ layer that the sandwich structure leads to frustration of exchange interactions. 

The magnetic moments in the FM and AFM state are presented in table \ref{Tab: MM}.
We observe two major trends: (i) the magnetic moments of Fe increase with the 
\textit{d}-band filling of the 5\textit{d} element and (ii) reducing the coordination number of Fe,
i.e.~if Fe is the top layer, gives rise enhanced magnetic moments.
Layers which are adjacent to the antiferromagnetic Fe are not spin-polarized due to the symmetry 
of the c(\(2 \times 2\)) AFM state.

The effects of hybridization at the interfaces and of the $5d$ band filling 
are visible in the local density of states shown in Fig.~\ref{Abb: LDOS} for the c$(2 \times 2)$ AFM state. If Fe is at the surface [Fig.~\ref{Abb: LDOS}(a)]
the LDOS is mainly influenced by the underlying $5d$ layer. Both the majority and the minority spin LDOS become sharper with
increasing $5d$ band filling. The hybridization in both channels, in particular, in the vicinity of the Fermi level is
also apparent. 

If the Fe layer is in the sandwich structure [Fig.~\ref{Abb: LDOS}(b)] the band width of both spin channels
increases due to the lower coordination and additional hybridization with the Rh surface layer. We observe an increased
majority LDOS above the Fermi energy and that the peaks in the minority spin channel are shifted above the Fermi level. 
The location at the surface leads to a reduced band width in the $5d$ layer. Layers adjacent to the Fe layer exhibit the same 
LDOS for majority and minority spin channels indicating that they are non-spin-polarized due to the symmetry in the c$(2 \times 2)$ 
AFM state which we consider here. If Fe is adjacent to the isoelectronic transition metals Ir and Rh, there is a matching of $3d$-$4d$ and $3d$-$5d$ hybridization. 
We observe states which are hybridized through the entire trilayer composed of Ir, Fe and Rh, e.g.~just above the Fermi energy. 

The collinear magnetic calculations show that bilayers with Ir are promising candidates for noncollinear magnetic structures with antiferromagnetic 
NN exchange interaction, which is underlined by the energy difference of Fe/Ir(001)\cite{Hoffmann2015}
in Fig.~\ref{Abb: Kollineare_Energien}. Therefore, we will focus on systems with an Fe/Ir interface in the rest of the paper.

\subsection{Freestanding FeIr bilayer} \label{Kap: Bilage}

\begin{figure}
\includegraphics[scale=1]{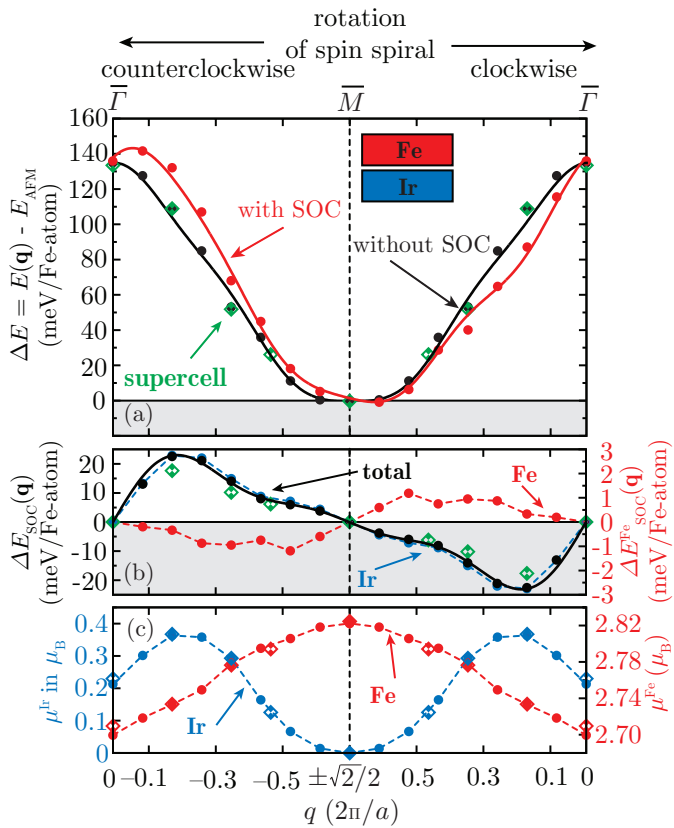}
\caption{(color online) (a) Calculated energy dispersion $E(\mathbf{q})$
of flat, cycloidal spin spirals
for a freestanding FeIr bilayer without (black dots) and with spin-orbit coupling 
(red dots) in \(\overline{M}-\overline{\Gamma}\)-direction with both senses of rotation.
The dispersion is fitted to the Heisenberg model (black line)
and including the DMI and magnetocrystalline anisotropy (red line).
The green diamonds indicate the values of the supercell calculations (see text for details).
(b) layer resolved contribution of \(\Delta E_{\text{SOC}}({\mathbf q})\).
The black curve is the fit of the DMI including three nearest neighbors.
(c) layer resolved magnetic moments.}
\label{Abb: FeIr}
\end{figure}

As a next step we isolate the FeIr interface and investigate an unsupported, freestanding FeIr bilayer
in view of noncollinear magnetic order. In Fig.~\ref{Abb: FeIr}(a) the energy dispersion $E({\mathbf q})$ 
of flat homogeneous spin spirals in the FeIr bilayer is shown along a high symmetry direction of the 2D BZ.
If we neglect SOC in our calculation clockwise- and counterclockwise-rotating spin spirals are energetically
degenerate. The lowest energy is obtained at the $\overline{M}$-point of the BZ which corresponds to the
c$(2\times 2)$ AFM state. The FM state ($\overline{\Gamma}$-point) is 138 meV/Fe-atom higher in energy exceeding the value found for 
FeIr bilayers on Rh(001) (cf.~Fig.~\ref{Abb: Kollineare_Energien}). From a fit to the Heisenberg model considering
up to 5th nearest neighbors we obtain the exchange constants given in table~\ref{Tab: FeIr Werte}. We find
a dominant NN interaction which is AFM (\(J_1 = -16 \, \text{meV}\)), however, exchange beyond NN is not
negligible. 

\begin{table}
 \centering
 \caption {Values of the $i$-th nearest neighbor exchange \(J_i\) (meV) 
 and Dzyaloshinskii-Moriya interaction constants \( D_i \) (meV) as well as the
 magnetocrystalline anisotropy (MAE) \(K \) (meV/Fe-atom) obtained for 
 the freestanding FeIr bilayer. 
 \(K < 0\) (\(K > 0\)) represents an out-of-plane (in-plane) easy axis.}
\begin{ruledtabular}
\begin{tabular}{cccccc}
\multicolumn{6}{c}{freestanding FeIr bilayer}  \\ \hline \hline
\(J_1\) & \(J_2\) &\(J_3 \) & \(J_4\) & \(J_5\) &  \\
\(-16.3\) & \(+3.1\) & \(-2.5\) & \(-0.3\) & \(-1.6\) &  \\ \hline
\( D_1\) & \( D_2  \) &\( D_3\) & \( D_4 \) & \(D_5\) & K \\
\(+5.7\) & \(-2.4\) & \(+4.5\) & \(+0.6\) & \(-0.7\) & \(-2.4\) \\
\end{tabular}
\end{ruledtabular}
\label{Tab: FeIr Werte}
\end{table}

Upon including SOC there is a preference for clockwise-rotating spin spirals and a small energy
minimum of 3.3 meV/Fe-atom occurs for a spin spiral period of 6.1 nm. Note that there is a small
shift of the energy dispersion of spin spirals with respect to the AFM state due to the 
magnetocrystalline anisotropy energy (MAE) which favors collinear states. In the AFM state the 
MAE favors an out-of-plane magnetization (cf.~table~\ref{Tab: FeIr Werte}).
The energy contribution due to 
SOC \(\Delta E_{\text{SOC}}({\mathbf q})\) has been obtained in first-order perturbation theory as discussed in 
section \ref{Kap: Methoden-DMI} and is displayed in Fig.~\ref{Abb: FeIr}(b).
We obtain maximal values of \(\Delta E_{\text{SOC}}({\mathbf q})\)  of more than 20 meV/Fe-atom. It
stems mainly from the Ir contribution due to its large SOC constant. In contrast the $3d$ 
transition metal Fe has a much smaller SOC constant and an almost negligible contribution. 
From a fit of $\Delta E_{\text{SOC}}({\mathbf q})$ we can obtain the strength of the DMI constants which are
given in table \ref{Tab: FeIr Werte}.
It is largest for the nearest neighbor (\(D_1 = 5.7 \,\text{meV/Fe-atom}\)) and exhibits an oscillatory character similar
to the exchange constants.
Due to the shape of $\Delta E_{\text{SOC}}$, e.g.~with different slopes at the \(\overline{\Gamma}\)
and \(\overline{M}\) point, it is necessary to include five nearest neighbors for the DMI fit.

\begin{figure}
\includegraphics[scale=1]{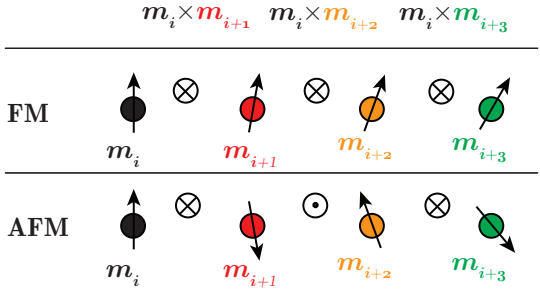}
\caption{One-dimensional sketch to illustrate the effect of DMI beyond nearest neighbors for 
clockwise rotating spin spirals close to the FM state (upper panel) and close to the AFM state (lower panel).
The cross product $\mathbf{m}_i \times \mathbf{m}_{j}$ is shown for the first three neighbors.
}
\label{Abb: DMI_AFM_Erklaerung}
\end{figure}

In Fig.~\ref{Abb: FeIr} (c) the magnetic moments of Fe and Ir layers are presented.
There is a small change of \(\mu^{\text{Fe}}\) and a strong spin polarization of Ir,
which has the same trend as \(\Delta E_{\text{SOC}}\). However, suppressing the spin 
polarization of Ir in the calculation by choosing a spin quantization axis perpendicular to that of Fe
gives rise to a very similar energy contribution due to SOC (see appendix).
Therefore, the DMI does not depend on the induced magnetic moment of Ir.

Close to the AFM state (\(\overline{M}\) point), the energy contribution due to DMI is reduced compared to that close to the FM state (\(\overline{\Gamma}\) point).
This is due to the competition of DM interactions beyond NN as apparent from the values and signs of the extracted 
DMI constants.
A one-dimensional example captures the essence of this effect as shown in Fig.~\ref{Abb: DMI_AFM_Erklaerung}.  
The first four spins of a clockwise rotating spin spiral along a chain of atoms are displayed. 
For the spin spiral with small angles between adjacent spins, i.e.~close to the FM state, the direction
of the cross product $(\mathbf{m}_i \times \mathbf{m}_j)$ which enters in the DMI term, Eq.~(\ref{Gl: DMI})
is always pointing into the page plane. 
Therefore, the energy due to DMI for $i$-th nearest neighbors will have the same sign if the DMI 
have the same sign. 
For a spin spiral in the vicinity of the AFM state 
(lower panel of Fig.~\ref{Abb: DMI_AFM_Erklaerung}),
on the other hand, the direction of the cross product between spins switches from one to the next
neighbor. Hence DM interactions with opposite signs would be favorable. 

Note that for a spin spiral along the \(\overline{\Gamma M}\) direction in the FeIr bilayer the spins on the 
second and third nearest neighbors possess the same canting angle $\phi={\mathbf q} {\mathbf R}_i$
(cf.~Fig.~\ref{Abb: Methoden-DMI}). Therefore, within the one-dimensional sketch they would both correspond
to the 2nd neighbor along the chain. From table \ref{Tab: FeIr Werte} we see that the sign of $D_1$, 
$D_2+\sqrt{2} D_3$ and $D_4$ are the same (the factor $\sqrt{2}$ results from evaluating the energy
for a cycloidal spin spiral along \(\overline{\Gamma M}\))). Therefore, we obtain a large energy contribution to the dispersion
of spin spirals close to the \(\overline{\Gamma}\) point (corresponding to the upper panel in Fig.~\ref{Abb: DMI_AFM_Erklaerung}) 
and a smaller one close to the \(\overline{M}\) point (lower panel in Fig.~\ref{Abb: DMI_AFM_Erklaerung}).

We expect a small error based on treating spin-orbit coupling in first order perturbation theory.
Therefore, we also perform self-consistent total energy calculations for spin spiral states 
in supercell geometries with and without SOC. We choose spin spiral states
with angles between the magnetic moments of adjacent Fe atoms
of \(0^\circ\) (\(\vert \mathbf{q}\vert = q = 0\)),
\(\phi=45^\circ\) (\(q \approx 0.18 \cdot \frac{2\pi}{a}\)),
\(\phi=90^\circ\) (\(q \approx 0.35 \cdot \frac{2\pi}{a}\)),
\(\phi=120^\circ\) (\(q \approx 0.47 \cdot \frac{2\pi}{a}\))
and \(\phi=180^\circ\) (\(q \approx \frac{\sqrt{2}}{2} \cdot \frac{2\pi}{a}\)).
The FM state (\(\phi=0^\circ\)) and the AFM state
(\(\phi=180^\circ\)) are calculated in each supercell geometry as a reference energy state.
The 2D unit cells corresponding to the spin spiral periodicities are
c(\(2 \times 8\)) for \(\phi=45^\circ\), i.e. 8 atoms per layer,
c(\(4 \times 4\)) for \(\phi=90^\circ\), i.e. 4 atoms per layer and
c(\(2 \times 6\)) for \(\phi=120^\circ\), i.e. 6 atoms per layer.

The diamonds in Fig.~\ref{Abb: FeIr} indicate the calculated total energies of these
states with respect to the AFM state neglecting SOC. The corresponding values are in 
very good agreement with the spin spiral calculations using the generalized Bloch theorem
(Fig.~\ref{Abb: FeIr}(a)). 
The values of the magnetic moments in Fig.~\ref{Abb: FeIr}(c) also match perfectly.
The only difference between both computational methods is in the contribution of SOC.
Indeed, the supercell calculation (green diamonds in Fig.~\ref{Abb: FeIr}(b))
show a similar trend of high values for the investigated states.
However, there is a slight energy difference which amounts to about 20\%.
We conclude that calculations of the SOC contributions to spin spiral states in first order perturbation
theory give the same trends and similar magnitude as self-consistent calculations.

\begin{figure}
\includegraphics[scale=0.95]{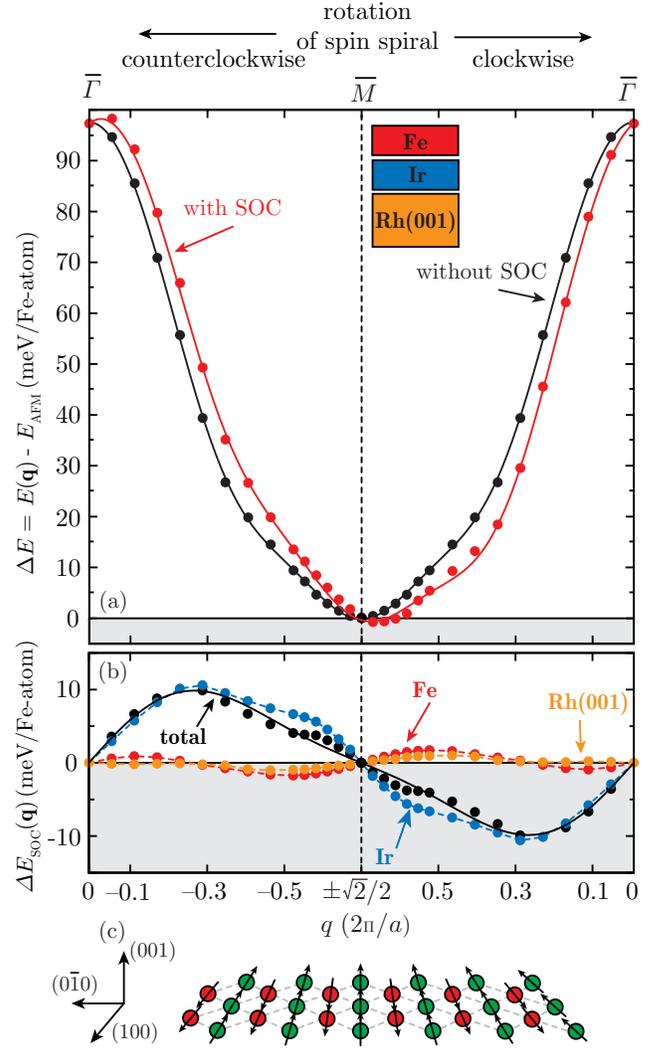}
\caption{(color online) (a) Calculated total energy dispersion $E(\mathbf{q})$
of flat, cycloidal spin spirals
for  Fe/Ir/Rh(001) without (black dots) and with spin-orbit coupling 
(red dots) in \(\overline{M}-\overline{\Gamma}\)-direction for both rotational senses.
The dispersion is fitted to the Heisenberg model (black line)
and including the DMI and magnetocrystalline anisotropy (red line).
(b) layer resolved contribution to $\Delta E_{SOC}(\mathbf{q})$.
The black curve is the fit of the DMI for three nearest neighbors.
(c) Sketch of the spin spiral state according to the energy minimum of the red curve of (a).}
\label{Abb: FeIrRh}
\end{figure}

The magnetocrystalline anisotropy energy (MAE) is calculated in the AFM state (See Sec.~\ref{Kap: Methoden-MAE}).
The FeIr bilayer prefers an out-of-plane magnetization with \(K = -2.4 \,\text{meV/Fe-atom}\).
We calculated the MAE also for a freestanding Fe monolayer (ML) in (001) geometry with the same in-plane 
lattice constant as for the bilayer. It also prefers a magnetization direction out-of-plane 
with \(K = -1.2\,\text{meV/Fe-atom}\) in the AFM state. Although the Ir is non-spin-polarized
in the AFM state of the Fe layer, the MAE is enhanced by a factor of two which we attribute to
the hybridization at the Fe-Ir interface and change of electronic structure.

\subsection{Noncollinear magnetism in FeIr bilayers on Rh(001)} \label{Kap: Systeme}

In the previous section we have seen that the freestanding FeIr bilayer exhibits strong 
antiferromagnetic exchange between nearest neighbors as well as large values of the DMI 
which extends beyond nearest neighbors. In this section we study how the Rh(001) surface 
affects these conclusions and in how far the stacking of the FeIr bilayer matters.

\begin{figure}
\includegraphics[scale=0.8]{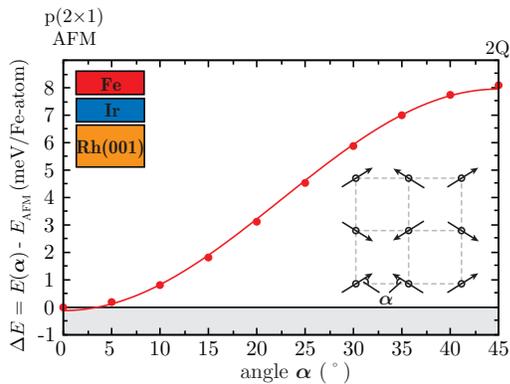}
\caption{(color online) Calculated energy of superposition states of spin spirals
for Fe/Ir/Rh(001) with respect to the p$(2 \times 1)$ AFM state. The considered spin 
structure is shown in the inset and \(\alpha\) is varied from 0 to \(45^\circ\). The
red line is a fit to the energy contribution for nearest neighbor biquadratic and
four-spin interaction (cf.~section \ref{Kap: Methoden-4Spin}). 
}
\label{Abb: FeIrRh 4-Spin}
\end{figure}

We start with the stacking in which the Fe layer is at the surface, i.e.~Fe/Ir/Rh(001). 
Figure \ref{Abb: FeIrRh}(a) shows the energy dispersion of flat spin spirals in Fe/Ir/Rh(001)
in \(\overline{M} - \overline{\Gamma}\)-direction.
The energy difference between the FM (\(\overline{\Gamma}\)) and c$(2\times 2)$ AFM (\(\overline{M}\)) state 
is similar to that of the collinear calculations (cf.~Fig.~\ref{Abb: Kollineare_Energien}) and to that 
reported for Fe/Ir(001) by Hoffmann \textit{et al.} \cite{Hoffmann2015}. 
From the energy dispersion without SOC we obtain the exchange constants given in table~\ref{Tab: FeIr auf Rh Werte}.
The exchange interactions between first (\(J_1 = -10.8 \,\text{meV}\)) and second nearest neighbors (\(J_2 = -3.8 \,\text{meV}\)) 
both try to align these spins antiparallel which is incompatible and leads to frustration. However,
the energetically lowest spin spiral state neglecting SOC is still at the \(\overline{M}\) point,
i.e.~the c$(2\times 2)$ AFM state.
The exchange constants are similar to those for Fe/Ir(001)~\cite{Hoffmann2015},
but differ considerably from those of the freestanding bilayer (cf.~Sec.~\ref{Kap: Bilage})

Upon including SOC, the Dzyaloshinskii-Moriya interaction (DMI) arises which leads to an energy
minimum in the spin spiral dispersion for clockwise rotating cycloidal spirals (see Fig.~\ref{Abb: FeIrRh}(a)) 
with an angle of about \(172^\circ\) from one to the next atomic row (see Fig.~\ref{Abb: FeIrRh}(c)).
The period of this spiral is about \(\lambda = 12 \, \text{nm}\).
Note, that the spin spiral energy curve with SOC in Fig.~\ref{Abb: FeIrRh} (a)
has been shifted by \(K/2 = 0.1\,\text{meV/Fe-atom}\) with respect to the c$(2 \times 2)$ AFM state.

The maximum energy contribution due to SOC amounts to \(10 \,\text{meV/Fe-atom}\).
It is mostly induced by the Ir layer with minor contributions from Fe and the Rh surface
as expected due to the large SOC constant of Ir. 
The large energy contribution due to SOC originates from the hybridization at the Fe-Ir interface.
The strength of the DMI can be seen in table \ref{Tab: FeIr auf Rh Werte}.
The DMI gains \(3.2 \, \text{meV}\) for the nearest neighbor while 2nd and 3rd neighbor contributions
are an order of magnitude smaller.

\begin{figure}
\includegraphics[scale=1]{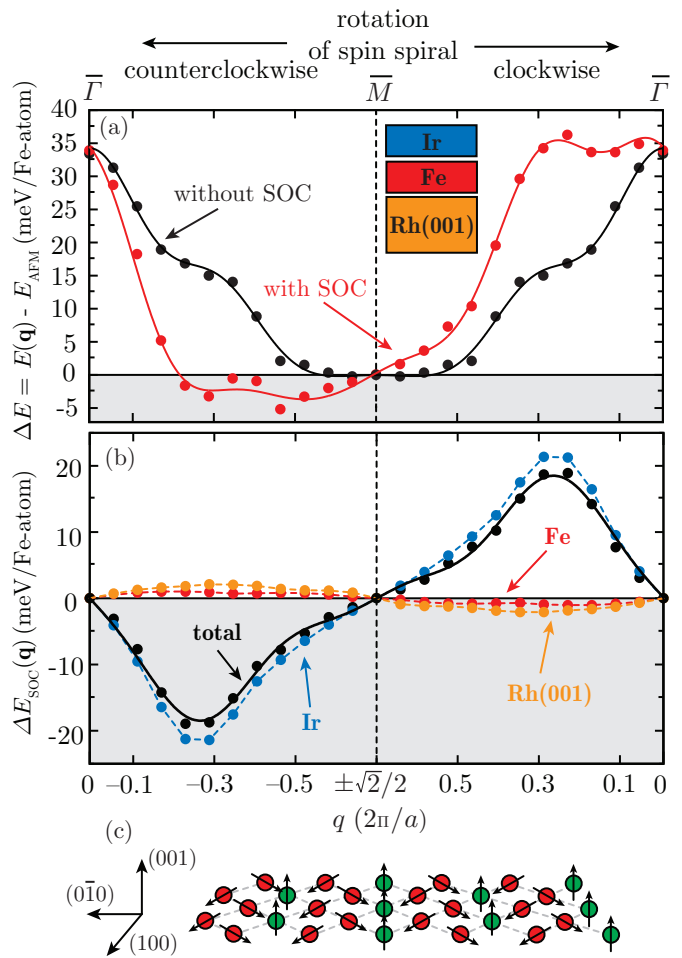}
\caption{(color online) (a) Calculated energy dispersion $E(\mathbf{q})$ of flat, cycloidal spin spirals
for Ir/Fe/Rh(001) without (black dots) and with spin-orbit coupling 
(red dots) in \(\overline{M}-\overline{\Gamma}\)-direction for both senses of rotation.
The dispersion is fitted to the Heisenberg model (black line)
and including the DMI and magnetocrystalline anisotropy energy (red line).
(b) layer resolved contributions to $\Delta E_{SOC}$.
The black curve is the fit of the DMI for three nearest neighbors.
(c) Sketch of the spin spiral state according to the minimum of the red curve of (a).}
\label{Abb: IrFeRh}
\end{figure}

The magnetocrystalline anisotropy energy (MAE) is \(K = +0.2 \, \text{meV/Fe-atom}\)
and prefers the spins to be in the plane of the film (cf.~Tab.~\ref{Tab: FeIr auf Rh Werte}).
It is interesting to compare the MAE to that of Fe monolayers on Ir(001) and Rh(001). 
While for Fe/Ir(001) a favorable out-of-plane magnetization has been found 
(\(K= - 0.25 \,\text{meV/Fe-atom}\) \cite{Hoffmann2015}
and \(K= - 0.56 \,\text{meV/Fe-atom}\) \cite{Belabbes2016}),
an easy in-plane magnetization axis was reported for Fe/Rh(001) \cite{Al-Zubi2011} 
(\(K= +0.2 \,\text{meV/Fe-atom}\)).
Surprisingly, the system Fe/Ir/Rh(001) behaves with respect to the MAE as Fe/Rh(001)
although the Ir layer is adjacent to the Fe layer. However, one has to remember that
we are considering the c$(2 \times 2)$ AFM state in which by symmetry the Ir layer
possess no induced spin polarization and only the Rh layer carries a magnetic moment
(cf.~table \ref{Tab: MM}). 

\begin{table*}
 \centering
  \caption {Values of the $i$-th neighbor exchange \(J_i\) (meV)
  and Dzyaloshinskii-Moriya interaction constants \( D_i \) (meV) as well as the
 magnetocrystalline anisotropy (MAE) (meV/Fe-atom) and higher-order exchange
 interactions for both stackings of the FeIr bilayer on Rh(001). 
 Note that we need seven neighbors for Fe/Ir/Rh(001) and 
 nine neighbors for Ir/Fe/Rh(001) to achieve a good fit for
 the exchange and three neighbors for the DMI for both systems.
 \(K < 0\) (\(K > 0\)) represents an out-of-plane (in-plane) easy magnetization axis.}
 \begin{ruledtabular}
\begin{tabular}{cccccccccccccccc} 
\multicolumn{5}{c}{Fe/Ir/Rh(001)} & & & \multicolumn{9}{c}{Ir/Fe/Rh(001)} \\ \hline \hline
\(J_1\) & \(J_2 \) & \(J_3\) & \(J_4\) & \(J_5 \) & & & 
\(J_1\) & \(J_2 \) & \(J_3\) & \(J_4\) & \(J_5 \) & \(J_6\) & \(J_7\) & \(J_8\) & \(J_9\)  \\
\(-10.8\) & \(-3.8\) & \(-0.7\) & \(-0.7\) & \(+0.4\) & & &
\(-3.4\) & \(+0.6\) & \(-0.8\) & \(-0.2\) & \(-2.3\) & \(-0.1\) & \(0.0\) & \(-0.2\) & \(+0.5\)  \\ \hline
\( D_1 \) & \( D_2 \) & \( D_3 \)
&\( K  \) & \(2K_{\text{4-spin}} + B \)
& & & \( D_1 \) & \( D_2 \) & \( D_3 \) & \(D_4 \) & \(D_5\) 
&\( K  \) &\multicolumn{3}{c}{\(2K_{\text{4-spin}} + B \)} \\
\(+3.2\) & \(+0.7\) & \(+0.3\) & \(+0.2\) & \(-2.0\)
& & & \(-5.3\) & \(+2.0\) & \(-2.9\) & \(+0.5\) & \(+1.2\) & \(-0.4\) &\multicolumn{3}{c}{\(-3.8\)}
\end{tabular}
\end{ruledtabular}
\label{Tab: FeIr auf Rh Werte}
\end{table*}

\begin{figure}
\includegraphics[scale=0.8]{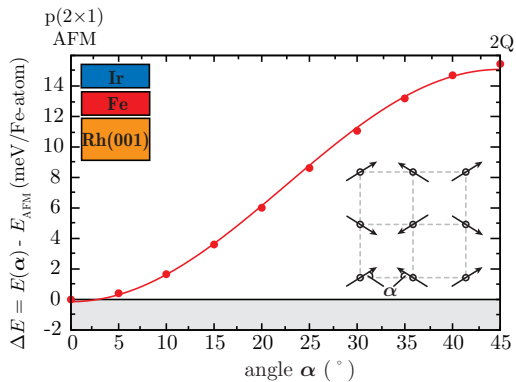}
\caption{(color online) Calculated energy of superposition states of spin spirals
for Ir/Fe/Rh(001) with respect to the p$(2 \times 1)$ AFM state. The considered spin 
structure is shown in the inset and \(\alpha\) is varied from 0 to \(45^\circ\). The
red line is a fit to the energy contribution for nearest neighbor biquadratic and
four-spin interaction (cf.~section \ref{Kap: Methoden-4Spin}). 
}
\label{Abb: IrFeRh 4-Spin}
\end{figure}

In strongly exchange-frustrated ultrathin film systems, it is possible that higher-order exchange interactions 
can compete with the Heisenberg exchange, DMI and MAE leading to complex magnetic ground states \cite{Heinze2011,Yoshida2012,Hoffmann2015}.
In order to estimate the importance of such terms in FeIr bilayers on Rh(001) we have calculated the total energy of superposition states 
of spin spirals as shown in the inset of Fig.~\ref{Abb: FeIrRh 4-Spin}. We vary the angle \(\alpha\) between \(0^\circ\), which corresponds to the
row-wise AFM state and \(45^\circ\), which is the so called 2Q-state \cite{Ferriani2007}.
These states are degenerate within the Heisenberg model, i.e.~there should be no change in energy with \(\alpha\).
However, in our DFT calculations we obtain an energy difference which is \(8 \,\text{meV}\) between \(\alpha=0^\circ\) and \(\alpha=45^\circ\)
which indicates the occurrence of higher order exchange interactions.
If we assume only nearest neighbor 4-spin and biquadratic interactions we expect the energy to vary
as $E(\alpha) = (2K_{\text{4-spin}} + B) \cdot \cos^2(2\alpha)$. As seen in Fig.~\ref{Abb: FeIrRh 4-Spin} we obtain an excellent fit to the
values from DFT resulting in \(2K_{\text{4-spin}} + B = -2\,\text{meV}\).
To determine the two constants separately further noncollinear spin states would have to be considered. Here we note that contributions from
higher-order interactions are of a similar order of magnitude as those from DMI.

The magnetic interactions presented above show similarities to those obtained in Fe/Ir(001)
where a spin lattice with AFM nearest-neighbor exchange interaction is predicted \cite{Hoffmann2015}.
Small deviations between the systems remain due to the different lattice constants
and the Rh vs.~Ir surface.
Additionally, the energy dispersion of Fe/Ir/Rh(001) around the AFM (\(\overline{M}\)) state
is similar to that of Pd/Fe/Ir(111) \cite{Dupe2014} close to the FM (\(\overline{\Gamma} \)) state
in which FM skyrmions could be observed expermentally \cite{Romming2013}.
In both systems there is a spin spiral minimum driven by the DMI resulting in a small canting
between adjacent spins with respect to the collinear state.
We conclude that Fe/Ir/Rh(001) is a promising ultrathin film system to find complex noncollinear 
spin structures such as AFM skyrmions or skyrmionic lattices 
with AFM nearest-neighbor exchange. 

Now we turn to the other stacking of the FeIr bilayer in which the Fe layer is
sandwiched between Ir and the Rh surface, i.e.~Ir/Fe/Rh(001).
The energy dispersion of spin spirals without SOC shown in Fig.~\ref{Abb: IrFeRh}
is in striking contrast to that of Fe/Ir/Rh(001). The energy difference between
the FM (\(\overline{\Gamma}\)-point) and the AFM (\(\overline{M}\)-point) state is smaller by more than a factor of two. The energy
dispersion is also extremely flat in the vicinity of the \(\overline{M}\)-point. 
As a consequence, we have to take into account more nearest neighbors
to obtain a reasonable fit to the Heisenberg exchange (black curve of Fig.~\ref{Abb: IrFeRh} (a)).
The obtained values of the exchange constants are given in table \ref{Tab: FeIr auf Rh Werte}.
The nearest-neighbor exchange is still antiferromagnetic but very small (\(J_1 = -3.6 \, \text{meV}\))
and exchange with further neighbors is of a similar magnitude. Hence there is a strong frustration of
exchange in this system.

The exchange frustration is also apparent upon including the energy contribution due to SOC
(Fig.~\ref{Abb: IrFeRh} (b)).
\(\Delta E_{\text{SOC}}(\mathbf{q})\) rises up to \(20\,\text{meV/Fe-atom}\), 
which is in the range of the total energy difference of 35 meV/Fe-atom
between the FM and the AFM state. Since Ir is on top of the Fe layer the DMI prefers left 
rotating cycloidal spin spirals in contrast to the right rotating spirals in freestanding FeIr bilayers
and in Fe/Ir/Rh(001). This change of the rotational sense is in accordance with the expectation from the 
Levy and Fert model \cite{Fert1980}.
The large maximum value as well as the shape
of \(\Delta E_{\text{SOC}}(q)\) is similar to that of the freestanding FeIr bilayer except for the opposite rotational sense (cf.~Fig.~\ref{Abb: FeIr}(b) ).
As expected, the main contribution stems from the Ir layer at the surface (see decomposition in Fig.~\ref{Abb: IrFeRh} (b)).

The DMI in Ir/Fe/Rh(001) is larger than the one of Fe/Ir/Rh(001) which is emphasized by the values of the DMI given in table \ref{Tab: FeIr auf Rh Werte}.
We obtain a DMI of \(5.3 \, \text{meV/Fe-atom}\) for the nearest neighbors that even exceeds the nearest neighbor Heisenberg exchange.
This has to our knowledge not been found for other systems so far. The values of the DMI are also large beyond nearest neighbors and
they are very similar to those found for the FeIr bilayer except for the sign due to the opposite rotational sense (cf.~table \ref{Tab: FeIr Werte}).
This shows the importance of the Ir layer being at the vacuum boundary with a reduced coordination and hybridization.
Another important difference to Fe/Ir/Rh(001) is that the MAE is \(K = -0.4 \,\text{meV/Fe-atom}\)
(cf.~Tab.~\ref{Tab: FeIr auf Rh Werte}), i.e.~preferring an out-of-plane magnetization.

Taking SOC into account we obtain quite a drastic change of the energy dispersion of spin spirals (Fig.~\ref{Abb: IrFeRh}(a)).
This is due to the large contribution from SOC as well as the strong exchange frustration in the
sandwich structure. The DMI leads to a canting of the spins into a spin spiral state with \(120^\circ\)
presented in Fig.~\ref{Abb: IrFeRh} (c).
Note that the fit to the dispersion is not perfect because deviations from fitting the 
exchange and DMI separately are summed up. The large values due to SOC obtained here
are similar to those of the freestanding FeIr bilayer which we confirmed by self-consistent
calculations (cf.~Fig.~\ref{Abb: FeIr}(b)). We conclude that changing the
stacking of the FeIr bilayer leads to a large enhancement of the DMI which we attribute
to the lower coordination and reduced band width of the Ir layer at the surface.

Higher-order exchange interactions may also play an important role to find the magnetic ground state in Ir/Fe/Rh(001).
As seen in Fig.~\ref{Abb: IrFeRh 4-Spin} the energy difference between the 2Q-state and the p$(2 \times 1)$ AFM state
has increased by almost a factor of two compared to Fe/Rh/Ir(001). The dependence of the energy on the angle $\alpha$
obtained from DFT is well described by considering nearest neighbor biquadratic and four-spin interaction leading to
a value of \(2K_{\text{4-spin}} + B =-3.8\)~meV.   
A non-vanishing biquadratic interaction would also affect the energy dispersion $E(\mathbf{q})$ of spin spirals while 
the four-spin term contributes only a constant energy shift.
The $\mathbf{q}$ dependence of the nearest neighbor biquadratic term is the same as that of the third nearest neighbor 
exchange interaction. Therefore, the fitting value given in table \ref{Tab: FeIr auf Rh Werte} for $J_3$ would then include 
the biquadratic term, i.e.~$ 2 J_3+ B=-0.8$~meV. Similarly, the second and third nearest neighbor biquadratic terms which
we expect to be even smaller would enter the fitting value obtained for $J_5$ and $J_9$. 
Due to the small values of the Heisenberg exchange
that are on the order of the DMI, higher-order terms should be able to compete and may become crucial for the magnetic 
ground state (cf.~Tab.~\ref{Tab: FeIr auf Rh Werte}). 
It will be a challenge for experimental studies to unravel the magnetic ground state of this system.

\subsection{Spin spiral calculations for Ir/Ir/Fe/Rh(001)} \label{Kap: IrIrFeRh(001)}

\begin{figure}
\includegraphics[scale=1]{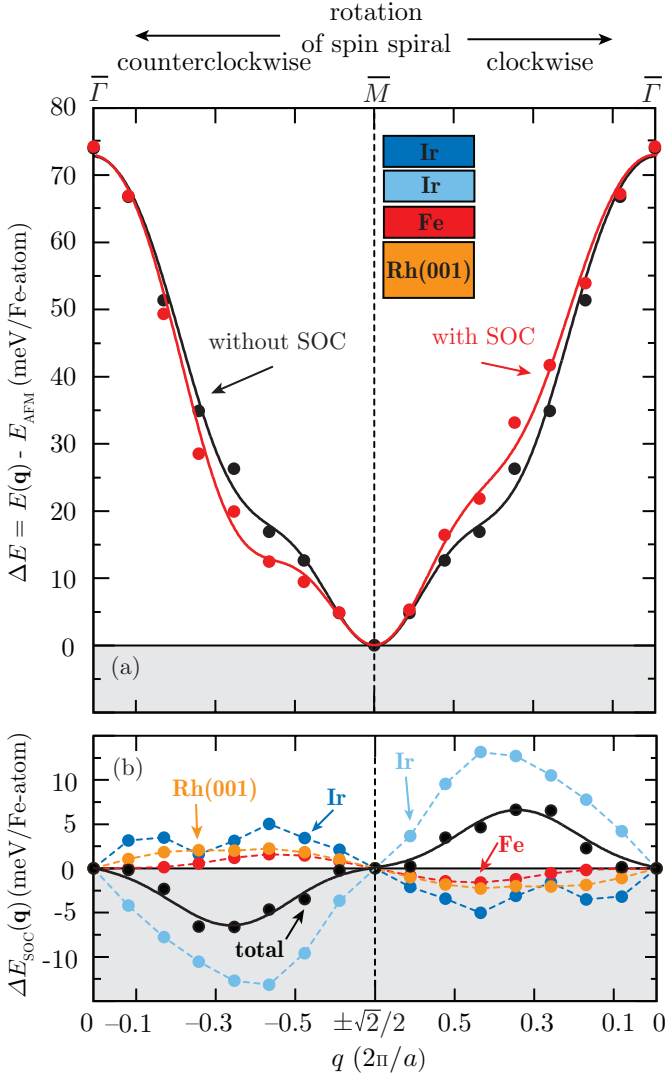}
\caption{(color online) (a) Calculated energy dispersion $E(\mathbf{q})$ of flat, cycloidal spin spirals
for Ir/Ir/Fe/Rh(001) without (black dots) and with spin-orbit coupling (red dots) in \(\overline{M}-\overline{\Gamma}\)-direction for both senses of rotation.
The dispersion is fitted to the Heisenberg model (black line)
and including the DMI and magnetocrystalline anisotropy (red line).
(b) layer resolved contribution to $\Delta E_{\rm SOC}(\mathbf{q})$.
The black curve is the fit of the DMI for three nearest neighbors.}
\label{Abb: IrIrFeRh(001)}
\end{figure}

Finally, we study the effect of an additional Ir adlayer on Ir/Fe/Rh(001) in order to
see whether the strong exchange frustration remains and whether the large DMI is an
effect of the lower coordination of Ir at the surface.
The energy dispersion of flat spin spirals without SOC is shown in Fig.~\ref{Abb: IrIrFeRh(001)}(a).
The total energy difference between the FM and c$(2 \times 2)$ AFM state is about \(75\,\text{meV/Fe-atom}\)
and the energy rises very fast close to the \(\overline{M}\)-point.
The obtained exchange constants are presented in table \ref{Tab: IrIrFeRh(001)}.
The nearest neighbor exchange rises by about a factor of two compared to Ir/Fe/Rh(001) and
becomes more dominant with respect to exchange beyond nearest neighbors. The exchange
frustration is thus reduced due to the additional Ir layer.

\begin{table}
 \centering
 \caption {Values of the $i$-th neighbor exchange \(J_i\) (meV) and
 Dzyaloshinskii-Moriya interaction constants \( D_i \) (meV) as well as the
 magnetocrystalline anisotropy (MAE) \( K \) (meV/Fe-atom) for Ir/Ir/Fe/Rh(001).
 All values are given in meV/Fe-atom.
 Note that we choose five neighbors for the exchange and 
 and three neighbors for the DMI for the fits.
 \(K < 0\) (\(K > 0\)) represents an out-of-plane (in-plane) easy magnetization axis.}
\begin{ruledtabular}
 \begin{tabular}{ccccc}
\multicolumn{5}{c}{Ir/Ir/Fe/Rh(001)}  \\ \hline \hline
\(J_1\) & \(J_2\) &\(J_3 \) & \(J_4\) & \(J_5\) \\
\(-7.3\) & \(-1.3\) & \(-1.0\) & \(-0.9\) & \(+1.0\) \\ \hline
\( D_1\) & \( D_2  \) &\( D_3\) & \(D_4\) & \( K \) \\
\(-2.0\) & \(+0.1\) & \(-0.2\) & \(+0.3\) & \(+0.4\)\\
\end{tabular}
\end{ruledtabular}
\label{Tab: IrIrFeRh(001)}
\end{table}

The energy contribution due to SOC $\Delta E_{\rm SOC}(\mathbf{q})$ reaches 
a maximum value of about \(6\,\text{meV/Fe-atom}\) (Fig.~\ref{Abb: IrIrFeRh(001)}(b))
and is much reduced compared to Ir/Fe/Rh(001).
There is still a very large contribution coming from the Ir-Fe interface with 
a value of up to \(15\,\text{meV/Fe-atom}\). However, it is balanced by the 
additional Ir layer and the Rh surface which act into the opposite direction. 
As a result the nearest neighbor DMI is reduced by about 60\% with respect to the IrFe bilayer
system (cf.~Tabs.~\ref{Tab: IrIrFeRh(001)} and \ref{Tab: FeIr auf Rh Werte}).

The magnetocrystalline anisotropy in the c$(2 \times 2)$ AFM state is reduced as well.
While we see an out-of-plane MAE in Ir/Fe/Rh(001) (\( K  = -0.4\,\text{meV/Fe-atom}\)),
it is in-plane upon adding an Ir adlayer \( K  = +0.4\,\text{meV/Fe-atom}\).
The Fe-Ir hybridization is weakened and thus the effect of the Rh substrate is 
intensified. While film systems with FeIr bilayers on Rh(001) are promising candidates for 
noncollinear magnetism with antiferromagnetic nearest-neighbor exchange interaction,
this is apparently not the case for the Ir/Ir/Fe trilayer on Rh(001).
The antiferromagnetic ground state driven by the exchange cannot be changed
because the DMI has a minor contribution to the total energy.

\section{Conclusions}\label{Kap: Vergleich}

We have studied the magnetic interactions in Fe/5\textit{d} bilayers on the Rh(001) surface
using density functional theory (DFT) as implemented in the FLAPW method. Upon changing the band
filling of the $5d$ transition metal from Os to Pt there is a transition of the nearest neighbor 
exchange interaction in the Fe layer from antiferro- to ferromagnetic. This effect occurs 
irrespective of the stacking of the bilayer, i.e.~with Fe at the surface or in the sandwich
geometry between the $5d$ layer and the Rh surface. However, in the sandwich geometry the nearest neighbor exchange is considerably
reduced which makes these systems prone to exchange frustration and complex ground states
due to competing interactions. 

In view of complex noncollinear magnetic states with antiferromagnetic nearest-neighbor exchange 
interaction such as isolated skyrmions and skyrmion lattices, we propose FeIr bilayers on Rh(001) as promising candidates. 
For both stackings of the bilayer we have obtained the exchange constants, the Dzyaloshinskii-Moriya 
interaction (DMI) and the magnetocrystalline anisotropy energy.
Higher-order exchange interactions are significant for both systems.

Fe/Ir/Rh(001) exhibits similar magnetic interactions as Fe/Ir(001) 
for which an atomic scale spin lattice has been predicted\cite{Hoffmann2015}.
However, the exchange and DMI differ slightly which may allow to find AFM skyrmions in this system.
It also has the advantages that
it is potentially easier to realize in experiments since Rh(001) does not possess a surface reconstruction 
and it allows fine tuning of the magnetic
interactions e.g.~by growing an additional Ir layer at the interface to the Rh surface.

Ir/Fe/Rh(001) is strongly exchange frustrated with very small values of the exchange constants.
The DMI is very large and even exceeds the Heisenberg exchange. DMI beyond nearest neighbors cannot 
be neglected.
We attribute the large values of the DMI 
in this system to the low coordination of the Ir layer at the surface. This is supported 
by similar values of the DMI for a freestanding FeIr bilayer. By including an additional Ir adlayer, on the 
other hand, the DMI is reduced to a much smaller value. The induced 
magnetic moment of the Ir layer does not affect the strength of the DMI.

\begin{acknowledgments}
It is our pleasure to thank
Gustav Bihlmayer and Matthias Bode for many fruitful discussions.
We gratefully acknowledge computing time at the supercomputer 
of the North-German Supercomputing Alliance (HLRN).
This project has received funding from the European Unions
Horizon 2020 research and innovation programme under grant agreement
No 665095 (FET-Opten project MAGicSky).
\end{acknowledgments}

\begin{appendix}

\section{Dependence of DMI on induced magnetic moments} \label{Kap: Anhang-MM Ir}

\begin{figure}
\includegraphics[scale=0.43]{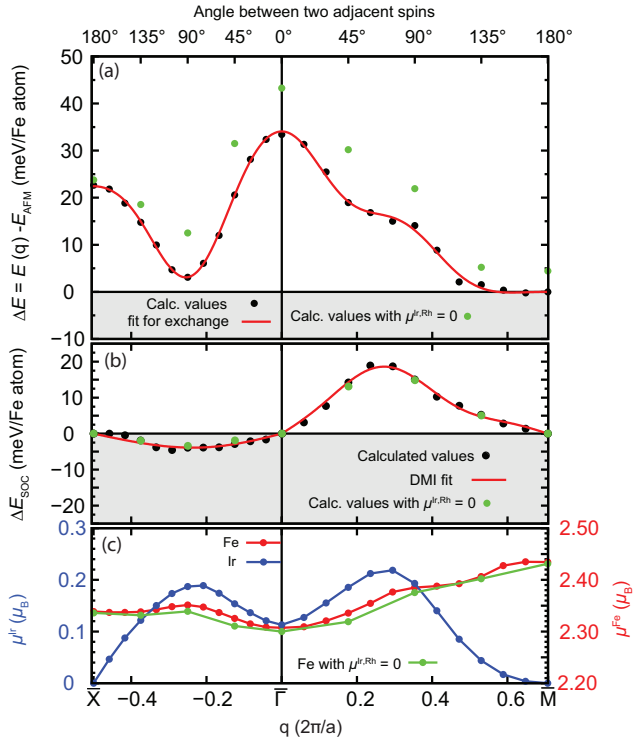}
\caption{(color online) Energy dispersion of spin spirals along the \(\overline{X}-\overline{\Gamma}-\overline{M}\)-direction
for Ir/Fe/Rh(001). (a) Energy dispersion $E(\mathbf{q})$ without spin-orbit coupling. (b) Energy contribution due to SOC, \(\Delta E_{\text{SOC}}(\mathbf{q})\) 
and (c) magnetic moments of the topmost three layers. The black points are the values including the induced magnetic moments in the Ir layer with the fit 
to the Heisenberg model and the DMI. The green points are values if the moments in the Ir layer are suppressed in the calculation.
}
\label{Abb: Spin-Polarisation}
\end{figure}

Figure \ref{Abb: Spin-Polarisation} shows the energy dispersion of spin spirals in Ir/Fe/Rh(001).
Compared to the figures in the main text, there are some differences.
We present the energy dispersion and $\Delta E_{\rm SOC}(\mathbf{q})$ along the \(\overline{X}-\overline{\Gamma}-\overline{M}\)-direction
which we also considered for all other systems in order to perform the fits to the Heisenberg model and the DMI.
The rotational sense is right rotating along \(\overline{X \Gamma}\) and
left rotating for the \(\overline{\Gamma M}\)-direction indicated by negative and
positive values of $q$, respectively. 
We have suppressed the induced magnetic moment of the Ir and Rh layers (green points)
within one of the calculations by choosing a spin quantization axis orthogonal to that of the Fe layer. 
The qualitative behavior of the energy dispersion without SOC [Fig.~\ref{Abb: Spin-Polarisation}(a)] remains the same as well as
the values of $\Delta E_{\rm SOC}(\mathbf{q})$ obtained in this way [Fig.~\ref{Abb: Spin-Polarisation}(b)].
These calculations show that \(\Delta E_{\text{SOC}}\) and hence the DMI does not depend
on the induced spin-polarization of the Ir and the Rh layers.

\end{appendix}

\newpage
\bibliography{literature}

\end{document}